\documentclass[11pt]{article}
\usepackage{fullpage}

\newtheorem{theorem}{Theorem}

\newcommand{\polylog}{\ensuremath{\mathrm{polylog}}}

\begin{document}

\title{A fast and simple $O (z \log n)$-space index for finding\\
approximately longest common substrings}
\author{Nick Fagan, Jorge Hermo Gonz\'alez and Travis Gagie}
\maketitle

\begin{abstract}
We describe how, given a text $T [1..n]$ and a positive constant $\epsilon$, we can build a simple $O (z \log n)$-space index, where $z$ is the number of phrases in the LZ77 parse of $T$, such that later, given a pattern $P [1..m]$, in $O (m \log \log z + \polylog (m + z))$ time and with high probability we can find a substring of $P$ that occurs in $T$ and whose length is at least a $(1 - \epsilon)$-fraction of the length of a longest common substring of $P$ and $T$.
\end{abstract}

\section{Introduction}

Finding a longest common substring (LCS) of two strings $P [1..m]$ and $T [1..n]$ is a classic problem, first solved in optimal $O (m + n)$ time fifty years ago when Weiner~\cite{Wei73} showed how to build a suffix tree for $T$ in $O (n)$ time.  Since a suffix tree for $T$ takes $\Omega (n)$ space with a fairly large constant coefficient hidden in the asymptotic notation, however, researchers have shown how to solve the problem quickly with a suffix array for $T$ (still in $\Omega (n)$ space but with a smaller constant), an  augmented compressed suffix array or bidirectional FM-index for $T$ (in space bounded in terms of the empirical entropies of $T$), LZ77-indexes and grammar-based indexes for $T$ (in $O (z \log n)$ space, where $z$ is the number of phrases in the LZ77 parse of $T$) and, most recently, with an index based on the run-length compressed Burrows-Wheeler Transform of $T$ (in space proportional to the number $r$ of runs in transformed $T$ plus $O (z \log n)$).  We refer the reader to Navarro's text~cite{Nav16} for descriptions of all these data structures except the last, which Bannai, Gagie and I~\cite{BGI20} proposed and Rossi et al.~\cite{ROLGB22} implemented recently.

For highly repetitive datasets such as genomic databases, LZ77-indexes and grammar-based indexes are the smallest of these data structures, but the approaches to finding LCSs with them so far are either slow or complicated.  Gagie, Gawrychowski and Nekrich~\cite{GGN13} gave the first non-trivial solutions, using either $O (z \log n)$ space and $O (m \log^2 z)$ query time or $O (z (\log n + \log^2 z))$ space and $O (m \log (z) \log \log z)$ query time.  Abedin, Hooshmand, Ganguly and Thankachan~\cite{AHGT22} showed how to use $O (z \log n)$ space and $O (m \log (z) \log \log z)$ query time with high probability.  Gao~\cite{Gao22} showed how to compute the matching statistics of $P$ with respect to $T$ --- from which in $O (m)$ time we can compute the maximal exact matches (MEMs) of $P$ with respect to $T$, a longest of which is an LCS of $P$ and $T$ --- using either $O (\delta \log (n / \delta))$ space and $O (m^2 \log^\varepsilon \gamma + m \log n)$ query time, or $O (\delta \log (n / \delta) + \gamma \log \gamma)$ space and $O (m^2 + m \log (\gamma) \log \log \gamma + m \log n)$ query time, where $\delta \leq \gamma \leq z$ are more sophisticated measures of compressibility.  Most recently, Navarro~\cite{Nav22} showed how to compute the MEMs of $P$ with respect to $T$ using $O (\delta \log (n / \delta))$ space and $O (m \log (m) (\log m + \log^\varepsilon n))$.  None of these solutions have been implemented, however.

In this paper we describe how, given $T$ and a positive constant $\epsilon$, we can build a simple $O (z \log n)$-space index such that later, given $P$, in $O (m \log \log z + \polylog (m + z))$ time and with high probability we can find a substring of $P$ that occurs in $T$ and whose length is at least a $(1 - \epsilon)$-fraction of the length of an LCS of $P$ and $T$.

\section{Data Structures}

Given a text $T [1..n]$ and a positive constant $\epsilon$, we first compute the LZ77 parse of $T$ and then build the sets
\[S_L = \left\{ T[i..j]\ :\ \begin{array}{l}
 \mbox{$j - i + 1 = \lceil (1 / (1 - \epsilon))^e \rceil \geq 1$ for some integer $e$ and} \\
 \mbox{some LZ77 phrase ends with $T [j]$}
\end{array} \right\}\]
and
\[S_R = \left\{ T[j + 1..k]\ :\ \begin{array}{l}
 \mbox{$k - j = \lceil (1 / (1 - \epsilon))^e \rceil$ for some integer $e$ and} \\
 \mbox{some LZ77 phrase begins with $T [j + 1]$}
\end{array} \right\}\]
of substrings of $T$.  Notice $S_R$ contains the empty substring but $S_L$ does not.

We compute the Karp-Rabin fingerprints of the $O (z \log n)$ substrings in $S_L$ and store those fingerprints in an $O (z \log n)$-space map that, given a fingerprint of a substring $T [i..j] \in S_L$, in constant time returns the co-lexicographic range of phrases ending with $T [i..j]$.  (Although it is the possible source of error in our result, for simplicity, we ignore the exponentially small probability of collisions.)  We also compute the Karp-Rabin fingerprints of the $O (z \log n)$ substrings in $S_R$ and store those fingerprints in an $O (z \log n)$-space map that, given a fingerprint of a substring $T [j + 1..k] \in S_R$, in constant time returns the lexicographic range of suffixes starting with $T [i..j]$ at phrase boundaries.  Finally, we build a $z \times z$ grid with a point at $(x, y)$ if the co-lexicographically $x$th phrase is immediately followed by the lexicographically $y$th suffix starting at a phrase boundary, and store an $O (z \log \log z)$-space data structure supporting $O (\log \log z)$-time 2-dimensional range-emptiness queries on this grid~\cite{CLP11}.

\section{Queries}

Farach and Thorup~\cite{FT98} observed that, by the definition of the LZ77 parse, the first occurrence of any substring of $T$ touches a phrase boundary (meaning the phrase boundary splits the substring into a non-empty prefix and possibly-empty suffix).  It follows that we can easily use our data structures as an $O (z \log n)$-space index for $T$ such that, given a pattern $P [1..m]$, in $O (m \log^2 (m) \log \log z)$ time we can find a substring of $P$ that occurs in $T$ and whose length is at least a $(1 - \epsilon)$-fraction of the length of an LCS of $P$ and $T$.  To do this, we first compute the Karp-Rabin fingerprint of each prefix of $P$, so we can compute the fingerprint of any substring of $P$ in constant time.  For each $j$ between 1 and $m$, each of the $O (\log m)$ values of $i$ such that $P [i..j]$ could be in $S_L$, and each of the $O (\log m)$ values of $k$ such that $P [j + 1..k]$ could be in $S_R$, we check whether $P [i..j] \in S_L$ and $P [j + 1..k] \in S_R$.  If so, we check whether there are any points on the grid in the rectangle that is the product of the horizontal co-lexicographic range of phrases ending with $P [i..j]$ and the vertical range of suffixes starting with $P [j + 1..k]$ at phrase boundaries.  If that rectangle is not empty, then $P [i..k]$ occurs in $T$.  This takes $O (\log \log z)$ time for each choice of $j$, $i$ and $k$, or $O (m \log^2 (m) \log \log z)$ time overall.

To see why we will find a substring of $P$ that occurs in $T$ and whose length is at least a $(1 - \epsilon)$-fraction of the length of an LCS of $P$ and $T$ (ignoring the possibility of collisions), let $T [i'..k']$ be the leftmost occurrence in $T$ of any LCS of $P$ and $T$.  By Farach and Thorup's observation, $T [i'..k']$ touches a phrase boundary; let $T [j]$ be the character immediately to the left of that phrase boundary, let $T [i]$ be the leftmost character with $i' \leq i$ such that $T [i..j] \in S_L$, and let $T [k]$ be the rightmost character with $k \leq k'$ such that $T [j + 1..k] \in S_R$ (remembering that $T [j+ 1..k]$ could be empty, in which case $k = j$).  Notice that $j - i + 1 = \lceil (1 / (1 - \epsilon))^e \rceil$ for some integer $e$ and $j - i' + 1 < (1 / (1 - \epsilon))^{e + 1}$ (otherwise we would choose a smaller value of $i$), so $(j - i + 1) > (1 - \epsilon) (j - i' + 1)$.  Similarly, $k - j = \lceil (1 / (1 - \epsilon))^e \rceil$ for some integer $e$ and $k' - j < (1 / (1 - \epsilon))^{e + 1}$ (otherwise we would choose a larger value of $k$), so $(k - j) > (1 - \epsilon) (k' - j)$.  Our search will find $T [i..k]$ (ignoring the possibility of collisions) and $k - i + 1 > (1 - \epsilon) (k' - i' + 1)$.

So far, however, our solution is slower than existing ones.  To speed it up, we keep track of the length $\ell$ of the longest match we have seen so far and, for each choice of positions $j$ and $i$ in $P$, we ignore choices of $k$ which cannot give us longer matches.  There are only $\polylog (m)$ possible lengths of matches we can return, so we spend a total of $\polylog (m + z)$ time checking matches for which we increase $\ell$.  It follows that we use $O (m \log (m) \log \log z + \polylog (m + z))$ time overall.

Our solution is still worse than Abedin et al.'s, so we now remove the $\log m$ factor in our query time.  Again, we ignore choices of $k$ which cannot give us longer matches than we have seen so far.  We also ignore matches for which we increase $\ell$, since we spend a total of $\polylog (m + z)$ time checking them.  Finally, we ignore choices of $j$, $i$ and $k$ such that $j - i + 1 < k - j$, since they are symmetric to cases in which $j - i + 1 \geq k - j$.  Therefore, for each choice of $j$, we consider only the $O (1)$ choices of $i$ such that $\ell / 2 \leq j - i + 1 \leq \ell$.  It follows that we use $O (m \log \log z + \polylog (m + z))$ time overall.  Apart from the $\polylog (m + z)$ term, this is the best query time achieved so far.

\begin{theorem}
Given a text $T [1..n]$ and a positive constant $\epsilon$, we can build a simple $O (z \log n)$-space index such that later, given a pattern $P [1..m]$, in $O (m \log \log z + \polylog (m + z))$ time and with high probability we can find a substring of $P$ that occurs in $T$ and whose length is at least a $(1 - \epsilon)$-fraction of the length of a longest common substring of $P$ and $T$.
\end{theorem}

\section{Future Work}

Although our index is simple and has competitive worst-case bounds, it is probably not practical.  We expect that we can make it practical, at the cost of worsening the worst-case bounds, by capping the maximum length of a pattern at $M$ --- so $S_L$ and $S_R$ have size $O (z \log M)$ --- and replacing the sophisticated range-emptiness data structure by a wavelet tree --- increasing the worst-case query time to $O (m \log z + \polylog (m + z))$ --- and replacing the LZ77 parse by the LZ-End parse~\cite{KN13} --- increasing the worst-case space bound to $O (z \log^2 n)$ according to current knowledge. 

We conjecture that switching to the LZ-End parse will actually reduce the space usage in practice, because no one has yet found a string on which the size of the LZ-End parse is $\omega (z)$ and, more importantly, if phrases' sources end at phrase boundaries then it seems likely many of the substrings we would add to $S_L$ and $S_R$ will be duplicates, which we can ignore.  We are currently implementing this modified solution and will report our experimental results in a future version of this paper.

\end{document}